\documentclass[showpacs,preprintnumbers,amsmath,amssymb,oneside]{revtex4}

\usepackage[dvips]{graphicx}
\usepackage{dcolumn}
\usepackage{bm}
\newcommand{\A}{a}

\begin{document}

\preprint{}

\title{Depth Development of Extensive Air Showers from Muon Time Distributions}

\author{L. Caz\'on}
\author{R.A. V\'azquez}
\author{E. Zas}
\affiliation{Departamento de F\'\i sica de Part\'\i culas, Facultade
de F\'\i sica,\\ Universidade de Santiago de Compostela, 15782
Santiago, SPAIN}


\begin{abstract}
We develop an algorithm that has the potential to relate 
the depth development of ultra high
energy extensive air showers and the time delay for individual muons. The time
distributions sampled at different positions at ground level by a large air
shower array are converted into distributions of production distances using an
approximate relation between production distance, transverse distance and time
delay. The method is naturally restricted to inclined showers where muons
dominate the signal at ground level but could be extended to vertical showers
provided that the detectors used can separate the muon signal from electrons
and photons. 
We explore the accuracy and practical uncertainties involved in the
proposed method. For practical purposes only the muons that fall outside the
central region of the shower can be used, and we establish cuts in transverse
distance.  The method is tested using simulated showers by 
comparing the production distance distributions obtained using the method with
the actual distances in the simulated showers. 
It could be applied in the search for  neutrinos to increase 
the acceptance to highly penetrating particles, as well as for unraveling 
the relative compositions of protons and heavy nuclei. We also illustrate that 
the obtained depth distributions have minimum width when both the arrival
direction and the core position are well reconstructed.  
\end{abstract}

\maketitle

\section{Introduction}
When an Ultra High Energy Cosmic Ray (UHECR) particle enters the atmosphere it
interacts producing an extensive air shower that propagates through it and
reaches ground level. These showers are routinely detected by optical systems
that collect fluorescence light emitted by nitrogen molecules excited as the
front crosses the atmosphere, and by arrays of particle detectors that sample
at ground level an enormous shower front which can exceed $10^{12}$
particles. In these arrays the relative times of the detected signals allow
the reconstruction of the incoming particle arrival direction.  The time
distribution of the arriving signal has been known for long to be dependent on
the depth distribution of the shower particles 
\cite{Lapikens,Linsley,Watson,Antoni:2002vv} 
which is different for different primary particles.

Exploring the highest energy particles is now considered to be a priority
because the data are scarce, there are discrepancies between results obtained
with the two techniques and because the origin of these particles is not at
all understood \cite{NaganoWatson}.  Their study is expected to provide both
information on violent objects in the Universe where these particles originate
and on their interactions (during propagation and in the atmosphere) at
energies exceeding those achieved in accelerator experiments by many orders of
magnitude.  New experiments are being constructed and devised to improve the
statistics, to increase the precision and to establish the mass of the primary
particles. The
Auger Observatory in Argentina is the first of a new generation of large
aperture experiments. It combines the two techniques and for the ground array
it uses water \v Cerenkov tanks with photomultiplier tubes and Flash Analog to 
Digital Converters (FADC) to record the time stamp of the signal in 25 ns 
intervals with unprecedented accuracy \cite{EA}.

The perspective of improved detectors has triggered an increase of
phenomenological activity in the study and characterization of extensive air
showers. Evaluation of the time structure of showers is part of this effort 
motivated by 
both the practical need of controlling the uncertainties in the arrival 
direction reconstruction and also by the hope that its understanding might 
shed new light on the challenging problem of establishing their 
composition. The idea of relating the muon distributions to the shower 
development has been already quite successful \cite{Danilova,Brancus:1997rr,Pentchev1,Ambrosio:1999nr,model}.
The arrival time distributions of muons has been characterized using
simple geometrical and kinematical arguments and making a key simplifying
assumption on the muon energy, transverse momentum, and distance of production
distributions, namely that they are independent \cite{cazon3}. Here we have
further developed the algorithm that relates the arrival time distribution of
muons to the depth development of the shower following on these ideas.

The scope of the method is limited because the shower front contains many 
other particles,
mainly electrons and photons.  In principle it requires muon identification
but fortunately the muon signal dominates in two circumstances: when the
shower is inclined and the electromagnetic part does not reach ground level
\cite{model} but also for ``more vertical'' showers when the distance to
shower axis is sufficiently large \cite{rhomu_rsignal}. The method complements
alternative depth reconstruction methods which are always limited when only
densities at ground level are taken into consideration. Moreover when the
effects of arrival angle and impact point misreconstruction are taken into
account it is seen that the induced distributions have a minimum in their
spread for the correct angle and impact point. This effect opens the
possibility of using this method for improving the confidence in the
conventional angular and impact point reconstructions. While the precision
obtained is possibly insufficient to be used for composition studies it will
certainly have an important impact on improving the acceptance of air shower
arrays to neutrinos through inclined showers. The accuracy in the depth
development reconstruction is sufficient to exclude neutrino interactions at
intermediate depths, when the electromagnetic shower would have been
completely absorbed but the first interaction is sufficiently deep into the
atmosphere to exclude both a cosmic ray hadron and a photon.

The article is organized as follows: In Section II we summarize the
factorization hypothesis for the muon distributions and the relations that
follow, and motivate the inversion of the relation between the time and depth
distribution from Ref.~\cite{cazon3}.  We also pay some attention to the
relation between particle densities and detector geometric acceptance. In
Section III we present the method to reconstruct the depth distribution. In
Section IV we compare the depth distributions obtained with this
reconstruction method to actual distributions from simulated showers to test
it. In Section V discussing some practical limitations. In Section VI we 
explore the correlations between the reconstruction procedure and the 
assumed arrival direction and impact point as a check of its robustness; 
we summarize and conclude in Section VII. Technical details are
presented in two appendices. 

\section{Relation between depth development and muon time distributions}

The main features of the arrival time distributions of muons in extensive air
showers can be accounted for by the different path lengths traveled by the
muons from their production point. This has been recently shown using a
simplified model to describe the muons in air showers which is based on the
hypothesis that their energy, $E_i$, transverse momentum, $p_t$, production
distance, $z$, and outgoing polar angle, $\zeta$ distributions factorize
\cite{cazon3}
\begin{equation}
\frac{d^4 N_0}{d z \;\! d\zeta \;\! d E_i \;\! d p_t
} = \frac{1}{2 \pi} {\cal N}_0 \;\!  h(z) \;\! f_1(E_i) \;\! f_2(p_t),
\label{factorization}
\end{equation}
In this expression $E_i$, $p_t$ and ${\cal N}_0$ refer to production, $z$ is
measured along the 
shower axis from the muon production point to the ground.  The
transverse momentum is transverse to shower axis and has polar angle $\zeta$
in the transverse plane (perpendicular to shower axis).  The functions $h(z)$,
$f_1(E_i)$ and $f_2(p_t)$ are assumed independent and normalized to 1 and the
factor $2 \pi$ accounts for a uniform polar angle distribution. Finally ${\cal
N}_0$, the total number of produced muons, is the overall normalization.  In
this model the muons are assumed to travel in straight lines and these four 
variables are sufficient to determine the muon path uniquely. 

It is convenient to express the
muon direction in terms of the angle $\alpha$ with respect to shower axis.
For a muon produced with energy $E_i$ and transverse momentum
$p_t$, the angle $\alpha$ is given by
\begin{equation}
\sin\alpha=\frac{p_t c}{\sqrt{E_i^2-(m c^2)^2}}\simeq\frac{ p_t c}{E_i},
\label{geometry_relation}
\end{equation}
We can approximate $p c = \sqrt{E_i^2-(m c^2)^2}\simeq E_i$, because the muon
energy at ground level is typically greater than $mc^2$, this energy at
production is even greater because of muon energy loss. 
We can now change the coordinates replacing $p_t$ in
Eq.~\ref{factorization} using Eq. \ref{geometry_relation} to give
\begin{equation}
\frac{d^4 N_0}{d z \;\! d\zeta \;\! d E_i
  \;\! d{\sin{\alpha}}} = {\cal N}_0 \;\! \frac{1}{2\pi}
h(z) \;\! f_1(E_i) \;\!  f_2\left(\frac{E_i}{c} \sin{\alpha} \right) 
\frac{E_i}{c}.
\label{fact_alpha}
\end{equation}
%
%
As the muons go through the atmosphere they lose energy and decay and 
even though we start with independent distributions at production, correlations
between the relevant variables appear naturally when we consider the surviving
muon distribution at ground level. This is explicitly shown in 
Appendix~A using a simplified model for energy loss. 

The distribution of surviving muons can then be integrated in $E_i$ in order to
obtain the depth distribution of the surviving muons, 
which is given in terms of
two angles describing the muon direction, namely $ \alpha$ and $ \zeta$. 
It is convenient to relate them to the differential solid angle for the muon 
$d^2 \Omega=-d\zeta d\cos \alpha$. Then
\begin{equation}
\frac{d^3 N}{d z \;\! d^2\Omega}=
\frac{d^3 N}{d z \;\! d\zeta \;\! d\sin{\alpha}}
\;\! \frac{\cos{\alpha}}{\sin{\alpha}}.
\label{dNdOmega}
\end{equation}

It is interesting to discuss in some detail the effect of a detector surface.
From Eq.~\ref{dNdOmega} we can obtain the number of muons from a given 
production interval $dz$ that crosses an arbitrary surface $d^2A$ which 
subtends a solid angle $d^2\Omega$: 
\begin{equation}
d N_A = \frac{d^3 N}{ d z \;\! d^2 \Omega} 
d^2 \Omega ~d z =  
\frac{d^3 N}{ d z \;\! d^2 \Omega} 
\;\! \frac{\cos{\psi_{\mu A}} \;\! d^2 A}{l^2} \;\! d z.
\label{eq:dN} 
\end{equation} 
where $ \psi_{\mu A}$ is the angle between the normal to
the surface and the muon direction.  On the other hand the projection of 
$d^2A$ onto the shower transverse plane is $r d \zeta d r= d^2 A \cos{\psi_A} $, 
where $\psi_A$ is the angle between the normal to the surface 
and the shower direction. (See Fig.~\ref{f:terra2} in Appendix A.) 
Using this relation we can relate Eq.~\ref{dNdOmega} to the number of 
particles per unit area in the transverse plane: 
\begin{equation}
\frac{1}{r} \;\! \frac{d^3N_A}{ dz  \;\! dr \;\! d \zeta} =
\frac{d^3 N}{ d z \;\! d^2 \Omega} 
\;\! \frac{1}{l^2} \;\! \frac{\cos{\psi_{\mu A}}}{\cos{\psi_{A}}} =
\frac{d^3 N}{ d z \;\! d^2 \Omega} 
\;\! \frac{1}{l^2} D_A(\Omega) =
\frac{d^3 N}{ d z \;\! d \zeta \;\! d\sin{\alpha}} 
\;\! \frac{\cos{\alpha}}{\sin{\alpha}} 
\;\! \frac{1}{l^2} D_A(\Omega).
\label{rho_mu} 
\end{equation}
Where $D_A(\Omega)$ denotes the geometrical factor involving
these two angles.  
We stress that 
$D_A$ not only depends on the surface orientation with respect to shower axis
but also on the direction of the incoming muons.  Any detector can be regarded
as a collection of such surfaces and as a result several such factors
$D_{A_i}$ will have to be considered depending on the 
impact point to obtain the
total effective collection area for a given arrival direction.

We can divide Eq.~\ref{rho_mu} by 
its integral in $z$ ($\hat N_{Ar \zeta}$) to normalize the function to 1. 
When this is done, using plausible functions for the distributions as 
described in Appendix A, 
and the relations between $r$, $l$, $z$ and $\alpha$ are used, a number of 
factors cancel out and we obtain the $z$-distribution of muons arriving at 
detector $A$ (normalized to 1) which can be related to a simple transform 
of the $z$ distribution: 
\begin{equation}
\frac{1}{{\hat N}_{Ar \zeta}} \;\! \frac{1}{r} \;\! \frac{d^3 N_A}{dr \;\! d z \;\! d\zeta} = 
\frac{h(z) \;\!  l^{1-\gamma} \cos \alpha ~D_A(\Omega)}{
    \int_0^{\infty} dz \;\! h(z) \;\! l^{1-\gamma} \cos \alpha ~D_A(\Omega)}.
\label{mainfrac} 
\end{equation}

The proposed method relies on the above expression and the geometrical 
relation between $z$ and $t$ which is described in Ref.~\cite{cazon3}. 
In that work it was shown that much of the time structure of
the muons is due to geometrical effects which imply that to each $z$ there
corresponds a given arrival time $t$. As a result we can relate the $t$ and
$z$-distributions:
\begin{equation}
\frac{1}{r}\;\! \frac{d^3 N_A}{dr \;\! d\zeta \;\! dt} =
\frac{1}{r} \;\! \frac{d^3 N_A}{dr \;\! d\zeta \;\! dz } \;\! \frac{dz}{dt}.
\label{dN_rdt} 
\end{equation}
We now define the normalized function $g(t)$ describing the shape of the time
distribution through:
\begin{equation}
{\hat N}_{Ar \zeta} \;\! g(t) = 
\frac{1}{r}\;\! \frac{d^3 N_A}{dr \;\! d\zeta \;\! dz} \Big| \;\! \frac{dz}{dt} \Big|.
\label{g(t)} 
\end{equation}
We can compare the $t$-distributions of the muons arriving at ground level
given by Eq.~\ref{g(t)} to those obtained in simulations and agreement is
found as will be shown in Section IV. 

The time distribution of the muons is related to the depth distribution of
muon production.  If we combine Eq.~\ref{mainfrac} and Eq.~\ref{g(t)} we
obtain the following relation between $h(z)$ and the time distribution:
\begin{equation}
g(t) \Big| \frac{dt}{dz} \Big| = 
\frac{1}{{\hat N}_{Ar \zeta}} \;\! \frac{1}{r} \;\! \frac{d^3 N_A}{dr \;\! d z \;\! d\zeta} = 
\frac{h(z) \;\!  l^{1-\gamma} \cos \alpha ~D_A(\Omega)}{
\int_0^{\infty} dz \;\! h(z) \;\! l^{1-\gamma} \cos \alpha ~D_A(\Omega)}.
\label{g(t)2} 
\end{equation}
This expression takes into account the fact that from different $r$ we
effectively sample the $h(z)$ distribution with an extra $z$-dependence
introduced as a overall factor through $l$ and the angles $\alpha$ and
$\zeta$.

\section{Reconstruction of the depth distribution}

In a typical air shower array a number of particle detectors sample the shower
front at the Earth's surface.  We now consider a set of $M$ detectors labeled
by a suffix $i$ (from $1$ up to $M$) each with a surface $A_i$ (which becomes
$S_i$ when projected onto the shower plane) and located at position
$(r,\zeta)_i$ in transverse plane coordinates.  We calculate the time
distribution of arriving muons to a detector $i$ by integrating 
Eq.~\ref{dN_rdt}
over the transverse surface $dS$, or, for all practical purposes, simply
multiplying by the corresponding area $S_i$.  Using Eq.~\ref{g(t)} the time
distribution at detector $i$ becomes:
\begin{equation}
\frac{dN_A}{dt}\Big|_{i} \equiv \int_{S_i}
\frac{d^3N_{A_i}}{dt \;\!  r \;\! dr \;\! d\zeta}(r,\zeta) dS 
\simeq  S_i \hat{N}_{Ar\zeta} ~ g(t).
\end{equation} 

The number of muons falling in the detector $i$ can be considered as a finite
sample of the continuous arrival time distribution probability
$\frac{dN_A}{dt}\Big|_i$. Let us assume that we can fill a time histogram with
$N_i$ entries corresponding to the $N_i$ muons detected by detector $i$.  The
entries of this histogram can be transformed into a $z$ histogram, using the
correspondence $t\rightarrow z$ given by \cite{cazon3}:
\begin{equation}
z=\frac{1}{2}\left( \frac{r^2}{ct}-ct\right)+ \Delta,
\label{z_t_exact} 
\end{equation} 
which can, in most cases, be approximated by:
\begin{equation}
z \simeq \frac{1}{2}\frac{r^2}{ct} + \Delta.
\label{z_t} 
\end{equation}
This mapping transforms each time entry 
(from $\frac{dN_A}{dt}\Big|_i \simeq S_i \hat{N}_{Ar\zeta} ~ g(t)$ ) 
into a $z$ entry (of $S_i \frac{1}{r}\;\! \frac{d^3N_A}{dr \;\! d\zeta \;\! dz }$) and 
finally into an entry of the $z$-distribution of the muons arriving at ground, 
$\frac{dN}{dz}$.

Note that the delay $t$ is the time difference between the arrival time of a
given particle and the arrival time of a reference plane perpendicular to the
shower axis and traveling at the speed of light $c$, the {\it time-reference
plane}. We have chosen the $0$-time origin corresponding to the arrival of the
first particle at ground at $r=0$ (shower core).  If the core hits ground at a
universal time $ct_0'$, the relation between $t'$ and $t$ involves $\Delta$:
\begin{equation}
ct=ct'-ct_0'+ \Delta.
\label{times_relation}
\end{equation}

Different detectors will give entries to a different time distribution, but
they will be converted into samples of a {\sl unique} $\frac{dN}{dz}$
distribution. As a result the converted entries of available detectors can be
combined into a larger sample.  These entries are naturally weighted by the
number of muons detected at each detector.

In Ref.~\cite{cazon3} it was shown that there is an additional source of delay
for muons because of their sub-luminal velocities $t_{\varepsilon}$.  The
total time delay is obtained adding it to the delay given by Eq.~\ref{z_t}:
\begin{equation}
t \simeq t_g+t_{\varepsilon}.
\end{equation}
This delay is energy dependent and it is only dominant over the geometric
delay for muons close to shower axis. In inclined showers the final muon 
energy is much larger than the muon mass, $m c^2 \ll E_i-\rho a l$, and 
therefore: 
\begin{equation}
t_\epsilon \simeq \frac{1}{2}\;\! \frac{(m c^2)^2}{c \rho a }
\left[\frac{1}{E_i-\rho a l} -\frac{1}{E_i}\right].
\label{t:kinetic} 
\end{equation}

\begin{figure}[htb]
\begin{center}
\includegraphics[width=15 cm]{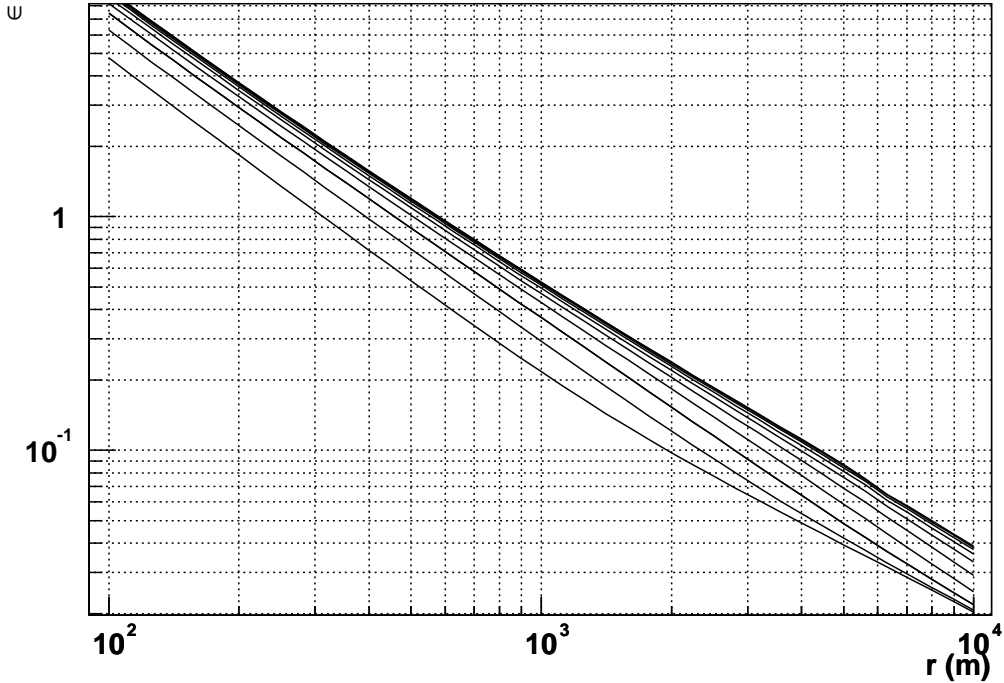}
\caption{The factor $ \epsilon(r,z)$ which relates to the ratio of kinematic
to geometric time delays (see text) versus $r$ for different $z$: 1.6, 3.2,
6.3, 13, 25, 50, 100, 200 km, from bottom to top.}
\label{f:epsilon_r}
\end{center} 
\end{figure}

Since the muon energy is not measured in typical air shower measurements, we
cannot account for these effects accurately. A solution is simply 
obtained by eliminating the measurements close to the axis to ensure that the
kinematical delay can be neglected. This has an impact on statistics.  
In Fig.~\ref{f:epsilon_r} we have plotted the factor $\epsilon (r,z)$, 
which can be taken as the relative value of the average kinematic delay with 
respect to the average geometric delay (see Appendix B), 
for $\zeta=90^{\circ}$, $\Delta=0$ and different $z$. We can see from this
graph that the geometric delay dominates for distances above 600~m from
shower core. 

We can include the kinematic delays on average. We obtain a simple
parameterization for the average kinematic delay as a function of $z$ and 
$r$, (details are given in Appendix~B):
\begin{equation}
\epsilon (r,z) = p_0(z)  \left(\frac{r}{\rm m}\right)^{p_1}.
\end{equation}
If we now subtract the average kinematical delay from the measured delay,
instead of Eq.~\ref{z_t_exact} we obtain: 
\begin{equation}
z \simeq \frac{1}{2} \left( \frac{r^2}{ct-c<t_{\varepsilon}>}
-(ct-c< t_{\varepsilon} >)\right)+\Delta.
\label{zcorr_t}
\end{equation}

Since it is not possible to obtain $z(t)$ analytically from the previous
expression we can use a simple numerical approach. We can for instance take
zero kinematical delay as a first approximation to obtain $z$, we then get the
average kinematical delay and substitute in Eq.~\ref{zcorr_t}. Since the
dependence of the $p$ coefficients on $z$ is mild (logarithmic) the procedure 
converges quickly and one iteration is sufficient.

\newpage

\section{Test}

One approach to test the method is to simulate showers using a
Monte Carlo
generator that reproduces the time distribution of the signal in a collection
of detectors and to apply the method to reconstruct the production
distribution of the muons $h(z)$. Unfortunately major modifications have to be
made in conventional simulation programs that are not designed to give the 
function $h(z)$ to compare with the reconstructed value.  We have
checked our reconstruction method by applying it to showers simulated with the 
Aires Monte Carlo package \cite{Aires}.  It is straightforward to compare the
$z$-distributions of the muons arriving at different locations in the 
ground. This has been done at different positions and agreement is found. 
This reflects the fact that the time distributions of the muons are well 
described by the model for muon time delays \cite{cazon3}. 
We have extended the test to combine detectors at different positions 
and to compare the results
to the total distribution of the surviving muons $dN/dz$, which is
straightforward to obtain in Aires and in most shower generators. In our model
this distribution would be obtained by integrating Eq.~\ref{mainfrac} over $r$
and $\zeta$ to cover all the ground:
\begin{equation}
\frac{dN}{dz}= \int \int \;\! \frac{1}{r}\;\! \frac{d^3N}{dr \;\!
  d\zeta \;\! dz} ~r \;\! dr \;\! d\zeta =
\int \int \hat{N}_{r\zeta} \frac{h(z) \;\! l^{1-\gamma}
\cos \alpha ~D_A(\alpha)}{ \int_0^{\infty} h(z) \;\! 
{l}^{1-\gamma} \cos \alpha ~D_A(\alpha) ~ dz} ~ r dr d\zeta.
\label{dNdz_integral}   
\end{equation}

Using simulations we have studied how the total $dN/dz$ distribution at
ground relates to the local distributions at different $r$, after integrating
in $\zeta$ and $r$. We first note that to a first approximation the
$\zeta$-integral is proportional to the integrand with $\Delta=0$. This is not
surprising since $\Delta$ changes sign when integrating over $\zeta$. We have
found that there is an effective value of $r$ ($r_*$) for local
distribution that gives a very
good approximation to the overall $dN/dz$ distribution. This value is
slightly zenith angle dependent and ranges from about 400~m for showers at
$0^\circ$, up to 1000~m for horizontal showers at $80^\circ$ or 1800~m at
$86^{\circ}$ . This is reasonable because the bulk of the muons arrive to
ground in a relatively constrained region: for instance, at $0^{\circ}$ this
region is between $\sim 60$~m and $\sim 1000$~m.  We can then substitute in
Eq.~\ref{dNdz_integral} $l$ for $l_{\star}=\sqrt{r_{\star}^2+z^2}$ and
$\alpha$ for $\alpha_{\star}=\arcsin\frac{r_{\star}}{l_{\star}}$ and also
consider that to compare with Aires that gives directly the muon
position it is not necessary to include a geometric correction factor,
i.e. $D_A=1$. We finally obtain the following approximation
\begin{equation}
\frac{dN}{dz} \propto  h(z) \;\! l_{\star}^{1-\gamma}
\cos \alpha_{\star} .
\label{dNdz_integral_aprox}
\end{equation}

Since in a practical air shower array the detectors are going to be arranged
on an unknown and arbitrary pattern around the shower axis it is convenient to
correct the $z$-distribution obtained at each detector to a common observation
distance $r_{\star}$ which approximately reproduces the overall $dN/dz$
as follows:

\begin{equation}
\frac{dN}{dz} \propto 
g(t) \Big|\frac{dt}{dz}\Big| ~ \times ~ \frac{ l_{\star}^{1-\gamma}
\cos \alpha_{\star} }{ l^{1-\gamma}  \cos \alpha} \frac{1}{~D_A(\alpha)}.
\label{geometric_correction}
\end{equation}
In general, we must divide by $D_A(\alpha)$ to remove the dependence on the
detector geometry if necessary.  We note that the correction factors $\frac{
l_{\star}^{1-\gamma} \cos \alpha_{\star}}{l^{1-\gamma} \cos \alpha}$
approach 1 when $z$ increases (i.e. in horizontal showers). Taking this into
account one can use a single $r_\star=400~$m for all zenith
angles and still obtain
relatively good approximations.

\begin{figure}[htb]
\begin{center}
\includegraphics[width=13 cm]{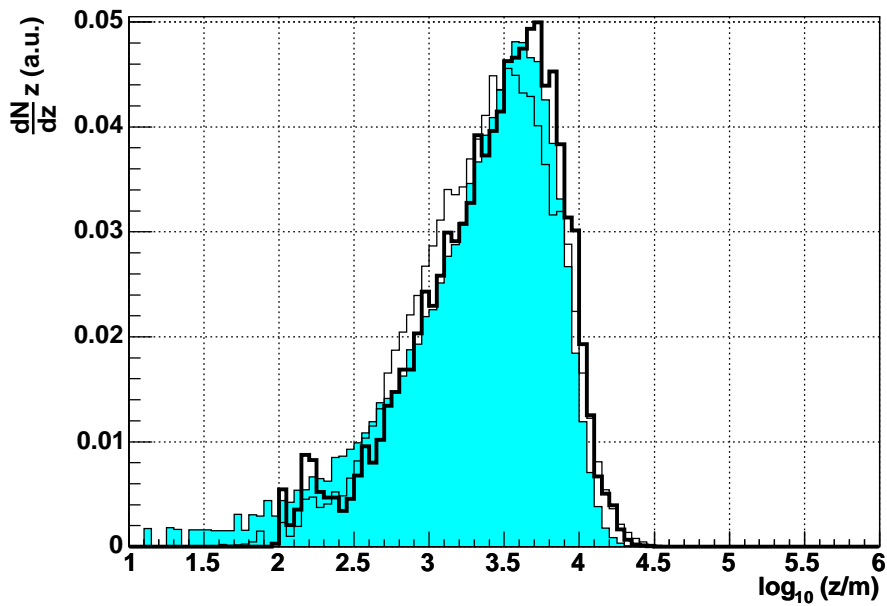}
\caption{Histograms of production distribution for 500 showers of $0^{\circ}$
zenith angle and $10^{19}$ eV energy: {\bf Light fill}: original
distribution. {\bf Unfilled Thick Line}: Final reconstruction, after all
corrections described in the text.  {\bf Unfilled Thin Line} Reconstruction
with no r cut. }
\label{f:00}
\end{center} 
\end{figure}
\begin{figure}[htb]
\begin{center}
\includegraphics[width=13 cm]{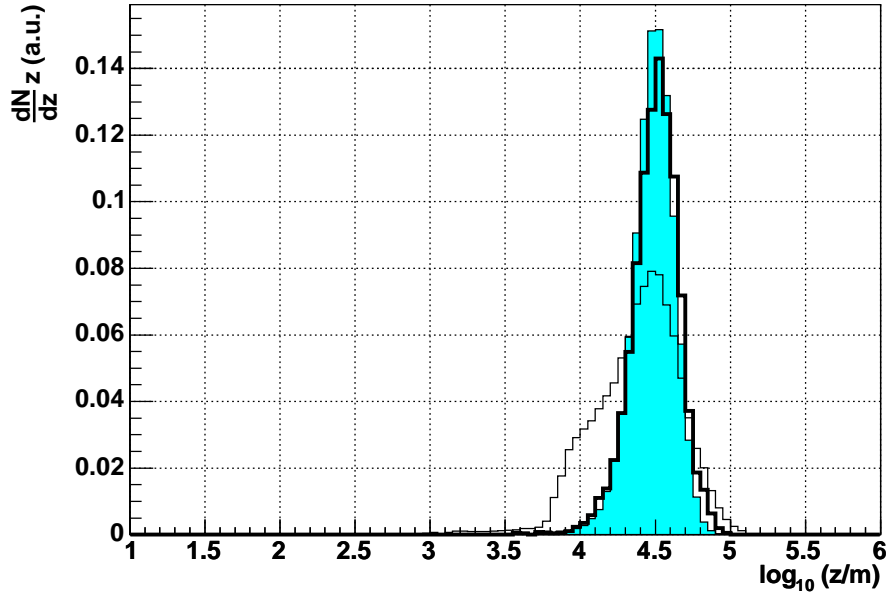}
\caption{Histograms of production distribution for 500 showers of $70^{\circ}$
zenith angle and $10^{19}$ eV energy: {\bf Light fill}: original
distribution.{\bf Unfilled Thick Line}: Final reconstruction, after all
corrections described in the text.  {\bf Unfilled Thin Line} Reconstruction
with no r cut. }
\label{f:70}
\end{center} 
\end{figure}

To test the method we have used sets of 500 showers at different zenith 
angles. 
A given particle array will be limited to a sample of this distribution which
is determined by the relative positions of the available detectors. 
We first take the muon output from the simulations and arrange the muons in 
a time histogram as can be done in an actual shower array (we use 25~ns bins). 
We then apply Eq.~\ref{zcorr_t} to all the simulated muons to calculate $z$,
where we have included the geometric corrections of 
Eq.~\ref{geometric_correction}
and the kinematic corrections as explained in the
previous section.
Finally an $r$ cut is applied.
In Figs.~\ref{f:00} and \ref{f:70} we illustrate the result for a $0^\circ$
and a $70^\circ$ showers. The 
shaded histogram is the distribution of all the muon production altitudes 
compared to that obtained from the 
reconstruction procedure using all the muons which reach the ground with 
$r>r_c$. This cut is necessary for the geometric inversion procedure to 
hold accurately. 
The result indicates that provided the muon time, the shower direction and
impact point coordinates are known, the reconstruction procedure works well. 
Figs.~\ref{f:00} and \ref{f:70} also show the same histogram without the $r$-cut which clearly fails to reproduce the $z$-distribution obtained in the 
simulation. This is mostly because of the time accuracy of the detectors
assumed to be $\sim 12.5~$ns. 


%

We have verified that the reconstructed histogram is not sensitive to small 
changes in the $r_c$ but clearly these cuts in $r$ can have large impact on 
the statistics. In table~\ref{t:shifts} we compare the averages of the ratio
of $z$ obtained with this method to that from the simulation $\Gamma=< \frac{z_{\rm rec}}{z_{\rm true}} >$. 
Clearly the method works best for moderately inclined showers between 
$60^\circ$ and $80^\circ$. 
At very low zenith angles, there is an overestimation of the production
distance, which could be due to an slight overestimation of the energy
loss. On the other hand, at very high zenith angles the magnetic field effects begin to be important and
the time geometric relation underestimates the production
distance. Nevertheless, in both cases, the precision obtained is quite good.

\begin{table}[htb]
\begin{tabular}{|c|c|c||c|c|}

\hline
\ \ $\theta \; \; ({\rm deg})$\ \ &\ \  $\phi$ (deg) \ \ &\ \ B \ \ & cut r(m)
& $\Gamma$ \\
\hline
 0  & -  &  no & 900 & 1.23 \\
30  & -  &  no & 1000 & 1.16 \\
60  & -  &  no & 1500 & 1.07 \\
70  & -  &  no & 2000 & 1.03 \\
80  & -  &  no & 2900 & 1.00 \\

80  & 0  & yes & 2900 & 0.97 \\
80  & 90 & yes & 2900 & 0.89 \\
86  & 0  & yes & 4000 & 0.86 \\
86  & 90 & yes & 4000 & 0.85 \\
\hline
\multicolumn{5}{l} {} \\
\hline
\multicolumn{5}{|c|} 
{\scriptsize neutrino-like  injected at $500$ ${\rm g/cm^2}$ vertical
  depth. } \\
\hline
80  & 0  & yes & 1400 & 1.02 \\
\hline
\multicolumn{5}{|c|} 
{\scriptsize neutrino-like  injected at $750$ ${\rm g/cm^2}$ vertical
  depth. } \\
\hline
80  & 0  & yes & 900 & 1.14 \\
\hline
\end{tabular}
\caption{Table of deviation of the reconstruction respect to the real
  distribution. $\Gamma=< \frac{z_{\rm rec}}{z_{\rm true}} >$. The azimuth
  angle $\phi$ is measured counterclockwise respect to the local magnetic 
north. For high zenith angles the results obtained with and without magnetic
  field are compared.}
\label{t:shifts}
\end{table}

\begin{figure}[htb]
\begin{center}
\includegraphics[width=13 cm]{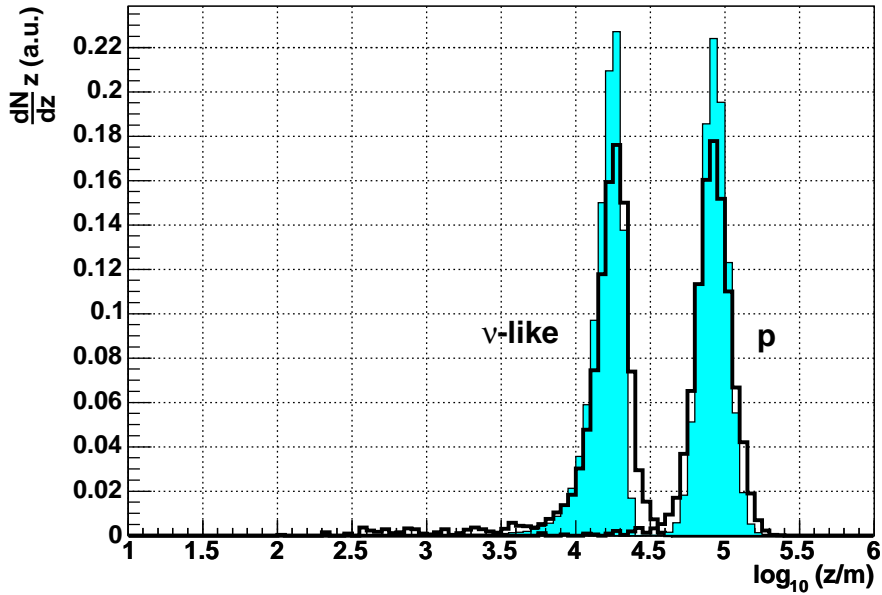}
\caption{Histograms of production distribution for 500 showers of $80^{\circ}$
zenith angle and $10^{19}$ eV energy for normal protons and for protons
injected at 500 ${\rm r/cm^2}$ of vertical depth to simulate a neutrino
interaction (marked as $\nu$-like). The geomagnetic field is included. 
{\bf Light fill}: original
distribution. {\bf Unfilled Thick Line}: Final reconstruction, after all
corrections described in the text.}
\label{f:p_and_nu500}
\end{center} 
\end{figure}

In Fig.~\ref{f:p_and_nu500} we have compared the results of the reconstruction
procedure applied to protons and deeply injected protons arriving with
$80^{\circ}$ zenith angle to illustrate how the method can be used to identify
deeply interacting inclined showers at high zenith, which are natural neutrino
candidates. A systematic study of the reconstruction procedure
  and the ability to identify neutrinos under realistic experimental
  conditions is left for future work.

\section{Limitations of the method}

The proposed method gives the reconstruction of the function $\frac{dN}{dz}$
i.e., the production altitude distribution for the surviving muons but clearly
the method is only valid with restrictions. Since the method addresses the
muon time distributions, it is essential to identify the muon contribution in
the shower front.  Clearly if the muon signal is separated at detector level,
there are no limitations due to this issue but this is in general not possible
for most of the detectors used in air shower arrays which typically have
complex signals containing a mixture of electron, positron and photon signals
(the electromagnetic component) and muons.

\v Cerenkov detectors, such as the water tanks used in Haverah Park \cite{HP}
and in the Auger observatory \cite{auger}, have some advantages in this
respect.  Since shower muons are penetrating particles the signal they give in
\v Cerenkov detectors is basically proportional to their track through the
detector, while the electromagnetic component typically gives a signal which
is simply proportional to the energy carried by photons, electrons and
positrons, because it is all absorbed in it.  As a result the volume of the
detector determines the ratio between the muon and electromagnetic signals to
a good extent. Large detectors give high muon contributions in spite of the
muons being a small fraction of all the particles in the shower front.
Moreover under some circumstances these large sharp pulses could in principle
be isolated and there are efforts in this line to separate individual muon
pulses from the time structure of the Auger tank signals
\cite{auger}.

In any case, since the muon lateral distribution is flatter than the
electromagnetic contribution \cite{ldfs}, the muons eventually dominate for
sufficiently large distances to shower axis.  As the zenith angle rises the
muons dominate closer and closer to shower axis. In close to horizontal
showers the muons dominate practically always~\cite{HP}.

Depending on the detector performance there are a number of 
limitations to the precision
which are addressed in this subsection.

It is firstly straightforward to see that the total number of entries of the
$dN/dz$ histogram is the total number of muons detected in all the
detectors, i.e $ N_{\mu}= \sum_{i=0}^M N_i$.  If we assume that the
$\frac{dN}{dz}$ distribution has a RMS width $\sigma$, this means that the
position of the mean $<z>$ (which is related to $X_{max}$) can be obtained
with no more precision than $\frac{\sigma}{\sqrt{N_\mu}}$ which is an
intrinsic statistical limitation.

A second limitation arises because of the intrinsic time resolution of the 
detectors used, $\delta t$, which will limit the precision of the muon 
arrival time. This will translate directly into an uncertainty in the 
production distance $z$, $\delta z$, through the map $t\rightarrow z$. 
According to Eq. \ref{z_t}
\begin{equation}
\frac{\delta z}{z} \simeq -\frac{c \delta t }{ct}
\left(1-\frac{\Delta}{z} \right)\simeq- \frac{\delta t}{t}.
\end{equation}
This equation can be rewritten to relate $\frac{\delta z}{z}$ to
$\delta t$ substituting $ct$ for the expression given by Eq.~\ref{z_t}:
\begin{equation}
\frac{\delta z}{z}=2 \frac{(z-\Delta)^2}{z}\frac{1}{r^2} c \delta t.
\label{eq:Dz_z}
\end{equation}

The time resolution of the detector affects the reconstruction precision
depending on distance to shower core. As we look at the arrival time of muons
closer to the shower axis the time delays become smaller and
the relative error on
the $z$ distribution reconstruction diverges. But again this problem can be
solved by imposing the cut $r>r_c$. To have $\frac{\delta z}{z}$ 
less than a given value $e_z$, and provided that we can find an approximated 
upper limit
for the production distance of the muons, $z<z_{u}$, we obtain the following
condition for the cut in $r$:
\begin{equation}
r > r_c(\zeta)=\frac{\sqrt{\frac{2  z_{u} c\delta t}{e_z}} }
{1+ \sqrt{\frac{2c \delta t}{e_z z_u}}{\tan \theta \cos \zeta}}.
\label{eq:r_c_a}
\end{equation}
Here $r_c$ depends on $\zeta$ because of the asymmetry induced by the term
$\Delta$. If we neglect this term, we obtain a simple expression that does not
depend on the angle $\zeta$:
\begin{equation}
r_c=\sqrt{\frac{2 z_{u} c \delta t}{e_z}}.
\label{eq:r_c}
\end{equation}

Notice that an $r$ cut can also avoid the regions near the shower axis where
the kinematical delay dominates over the geometrical, and also the region
where the muonic component signal is shadowed by the electromagnetic
component. In necessary case, the most stringent of the restrictions must be
applied.
 
\begin{figure}[htb]
\begin{center}
\includegraphics[width=10 cm]{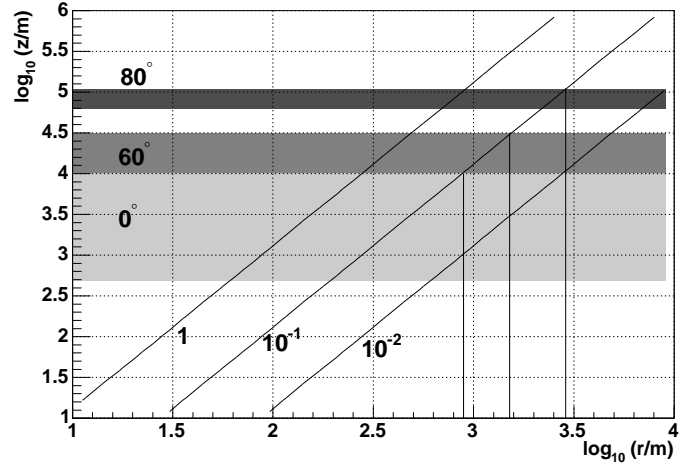}
\caption{Contour lines for the function $\frac{\delta z}{z}=2 \frac{z}{r^2} c
  \delta t$, with $\delta t= 12.5$ ns. Different grey bands 
  show the 1-$\sigma$ production
  distance of muons at different zenith angles.}
\label{Delta_z}
\end{center} 
\end{figure}

For example let us consider an air shower array with time resolution $\delta
t=12.5$~ns (corresponding to half the sampling rate of the Auger detector),
located at 1400~m altitude and detecting showers with a zenithal angle of
$60^{\circ}$. We can easily identify an upper limit for production distance 
(for instance using simulation), for $60^{\circ}$ it is $z_{u}=31.6$ km. 
According to Eq.~\ref{eq:r_c} we would
require that $r>1500$~m in order that the resolution on the
$z$-reconstruction, $\frac{\delta z}{z}$, was less that $10\%$ ($e_z=0.1$).
Fig.~\ref{Delta_z} illustrates the effect showing the contour lines of the
precision as a function of $r$ and $z$ as given by Eq.~\ref{eq:r_c}.  The
value of $r_c$ must increase as the zenith angle rises because the muons are
produced at higher $z$.  For high zenith the effect of the neglected
$\zeta$-dependent term makes the $z$-reconstruction somewhat worse than the
approximate expression of Eq.~\ref{eq:r_c}, but enough for our purposes (in
necessary case the full expression
(Eq.~\ref{eq:r_c_a}) could also be used).  For a shower with $\theta=86^\circ$
if we use $e_z=0.1$ with the former expression to obtain $r_c$ we actually get
a resolution $\frac{\delta z}{z}$ which rises up to $0.17$ for the worst case
corresponding to $\zeta=180^{\circ}$.

\section{Correlation with angular and core position uncertainties}

The method has a third intrinsic limitation because to convert the
arrival time histogram into a histogram of production distances using
Eq.~\ref{h_z} the incoming direction and the position of shower axis must be
known, so that the appropriate values of $r$ can be introduced. In a realistic
case these will only be known to a given precision and further uncertainties
will arise because of the correlations between both core position and
direction uncertainties with the distance distribution obtained. The study of 
these effects with simulations shows that the reconstructed 
distributions have a minimum in width when the true shower direction and
impact position are used to reconstruct the depth distribution. This adds a
interest to the method since it can be used in principle as a further check 
of the reconstructed directions and impact points. 

\begin{figure}[htb]
\begin{center}
\includegraphics[width=13 cm]{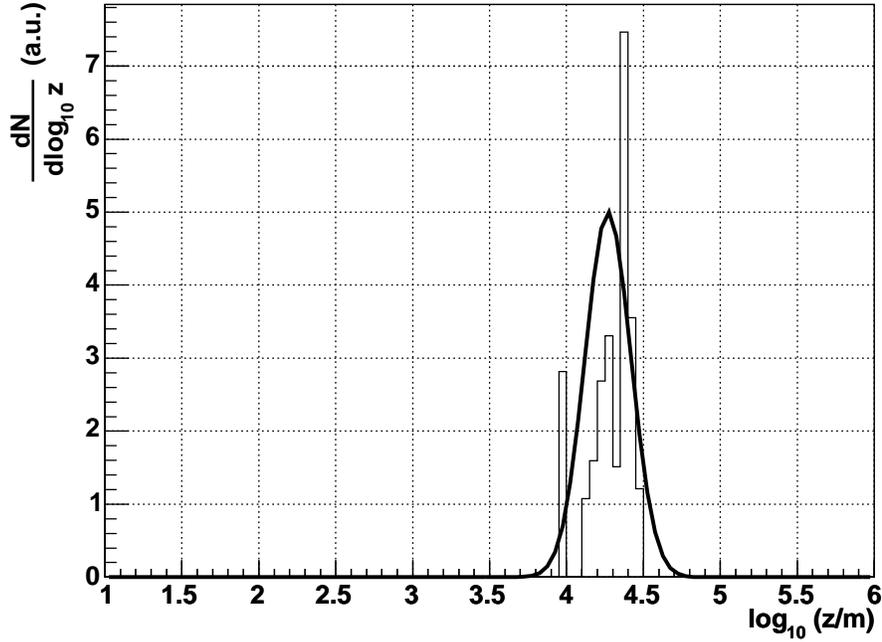}
\caption{Distribution for a reconstructed $10^{19}$ eV proton shower of
  $\theta=60^\circ$ and $\phi=90^\circ$ (histograms) compared with the
  original distribution when we have finite sampling produced by a finite number
  of detectors.}
\label{60_19dNdlogz_and_original}
\end{center} 
\end{figure}

\begin{figure}
\begin{center}
\includegraphics[width=7 cm]{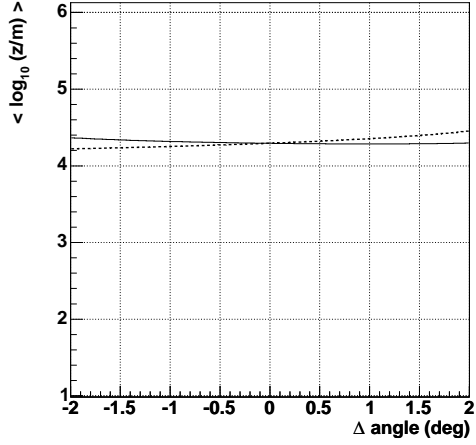}
\caption{Effects of reconstruction direction shifts for a $10^{19}$ eV proton 
shower of $\theta=60^\circ$ and $\phi=90^\circ$ with an $r$ cut $r>1500$ m. 
Mean of $\log_{10} (z/m)$ as a function of $\Delta \theta$ fixing $\Delta
\phi=0$ (dashed line) and as a function of $\Delta \phi$ fixing 
$\Delta \theta=0$ (continuous line).}
\label{mean60_19angle_1500}
\end{center} 
\end{figure}

\begin{figure}
\begin{center}
\includegraphics[width=14 cm,height=7cm]{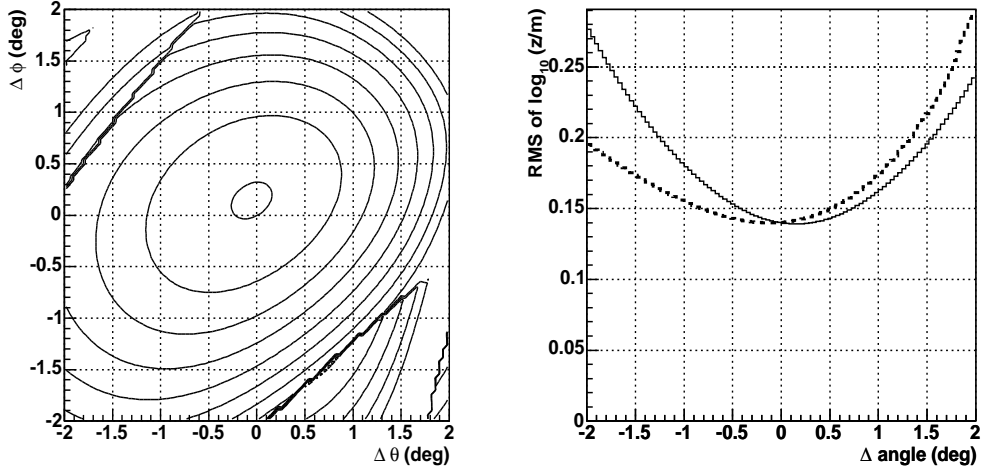}
\caption{Effects of reconstruction direction shifts in the width of the
$z$-distribution for a $10^{19}$ eV proton shower of $\theta=60^\circ$ and
$\phi=90^\circ$ with an $r$ cut $r>1500$ m.  {\bf Left} RMS width of
$\log_{10} (z/m)$ as a function of angular shifts $(\Delta \theta,\Delta
\phi)$. {\bf Right} RMS of $\log_{10} (z/m)$ as a function of $\Delta \theta$
fixing $\Delta \phi=0$ (dashed line) and as a function 
of $\Delta \phi$ fixing $\Delta \theta=0$ (continuous line).}
\label{RMS60_19angle_1500}
\end{center} 
\end{figure}

To study these correlations we explore the stability of the reconstruction
to shifts in the core positions and angular directions.  Unfortunately the
computing time necessary to test such stability can be very large if
simulations are used in the same way they were used to test the method in the
previous section. We will use instead the results of the muon time delay model
of Ref.~\cite{cazon3} to get distributions of the arrival time for the shower
muons from simulated showers.  In an attempt to be closer to experimental
conditions we assume an array of particle detectors and calculate the number
of muons that crosses each of them. We choose 10~$m^2~\times~1.2~m$ (area
$\times$ height) detectors in a hexagonal grid, separated 1500~m corresponding
to the Auger surface detector.  This limits the statistics of the
reconstructed distribution, $dN/dz$, in a realistic way.
Fig.~\ref{60_19dNdlogz_and_original} shows an example of the statistics that
could be obtained for a $10^{19}$~eV shower with $\theta=60^\circ$ and
$\phi=90^{\circ}$. ( The azimuth
  angle $\phi$ was measured counterclockwise respect to {\it East} direction.)
%
We first recalculate the depth distributions assuming that the incidence 
direction has been misreconstructed by $(\Delta \theta,\Delta \phi)$ with 
respect to the actual arrival direction chosen for the simulation. 
This procedure was repeated for different shifts in angular space  
within an interval of $4^\circ  \times 4^\circ$. 
For each angular shift both the mean and RMS width of the $z$-distributions in
log basis were calculated. 
In Figs.~\ref{mean60_19angle_1500} and~\ref{RMS60_19angle_1500} representative 
results showing the mean and RMS width for an example of a
$10^{19}$ eV proton shower of $60^\circ$ zenith are shown as a
function of $\Delta \theta$ ($\Delta \phi$) for fixed $\phi$ ($\theta$), 
Also the RMS width is shown in a two dimensional plot $(\Delta \theta, \Delta
\phi)$ in Fig.~\ref{RMS60_19angle_1500}. 
It is worth remarking that the mean value of $z$ is quite stable to shifts 
in azimuthal direction $\Delta \phi$, whereas there is a very slight rise of
the reconstructed $z$ when increasing the zenith angle $\theta$. The behavior
on this angle in not symmetrical since $\Delta$ introduces an asymmetry between
early and late regions.
On the other hand the RMS width of the distributions seems to have a local
minimum when the correct arrival direction is used. 
In a typical air shower array the arrival direction is obtained using the 
relative arrival times of the signals at different locations. 
The observed minimum of the RMS width of the $z$-distribution 
suggests that this method could be also used to either reconstruct shower 
direction independently or, more likely, to check that the 
reconstruction obtained through conventional methods is consistent with the 
arrival time of the muons at large distances from shower core, on an 
individual shower basis. 
%

\begin{figure}
\begin{center}
\includegraphics[width=15 cm,height=7cm]{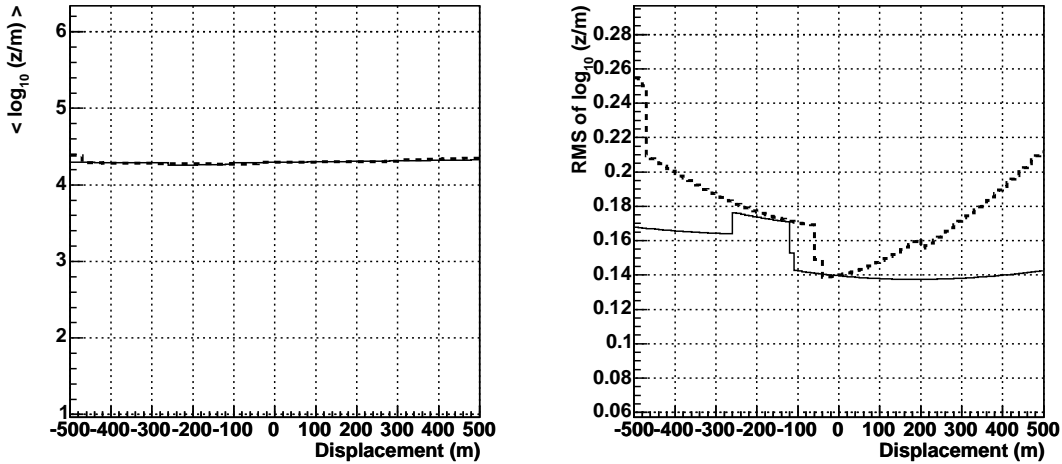}
\caption{Effects of impact point shifts on the mean (left panel) and RMS
  (right panel) of the $z$-distribution as a function of $\Delta x$ fixing
  $\Delta y=0$ (dashed line) and as a function of 
   $\Delta y$ fixing $\Delta
  x=0$ (continuous line) for a $10^{19}$ eV proton shower at 
  $\theta=60^{\circ}$ and
  $\phi=90^{\circ}$ with an $r$ cut $r>1500$ m.
\label{mean_and_RMS_60_19core_1500}
}
\end{center} 
\end{figure}

A completely analogous method was followed to study the core position and 
$z$-reconstruction correlations. The reconstructed impact points were shifted
by $(\Delta x,\Delta y)$ in the ground plane with respect to the core 
of the simulated shower over a grid covering a rectangle of 1000 
$\times$ 1000 m. The means and RMS widths for 
shifts in both $x$ and $y$ positions are shown in 
Fig.~\ref{mean_and_RMS_60_19core_1500}.
The plots display important discontinuities. 
They are of statistical nature because the total number of muons in the
detector is small and as the core position position is shifted individual 
detectors are rejected or accepted because of the $r$ cut. 
The detectors close to the cut are those that have most muons and including or 
not including them affects the $z$-distribution. These discontinuities are 
also present in some circumstances for angular shifts because the relative 
position of the detectors also change but clearly the changes of angular 
reconstruction modify the distances to shower axis at a second order level. 
These discontinuities
can become smoother by increasing statistics, for instance relaxing the $r$-cut.

\begin{figure}
\begin{center}
\includegraphics[width=14 cm,height=7cm]{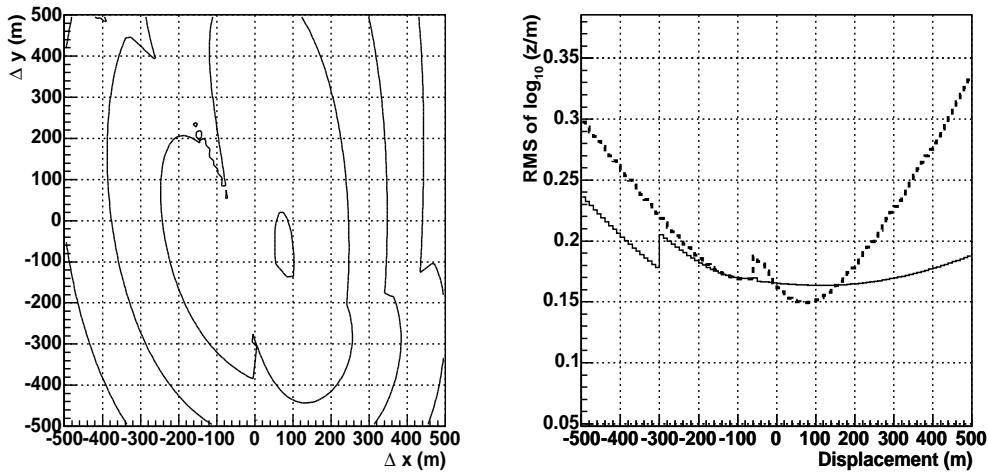}
\caption{Effects of impact point shifts on the RMS width of the
$z$-distribution for a $10^{19}$ eV proton at $\theta=60^{\circ}$ and
$\phi=90^{\circ}$ with an relaxed $r$ cut $r>500$ m .  {\bf Left} RMS of
$\log_{10} (z/m)$ as a function of a core shift $(\Delta x,\Delta y)$. {\bf
Right} RMS of $\log_{10} (z/m)$ as a function of $\Delta x$ 
(dashed line)
fixing $\Delta y=0$, and as a function of $\Delta y$ fixing $\Delta
x=0$ (continuous line).}
\label{RMS60_19core_500}
\end{center} 
\end{figure}

Notice that the mean value of $z$ is again quite stable to shifts in core
position.  This is not difficult to understand since each core location
corresponds to a new {\it time-reference plane} which is only slightly shifted
along the shower axis with respect to the planes obtained for other core
positions. The effect is due to the curvature of the front and is therefore a
second order effect.  The RMS width of the distributions also displays a local
minimum when the correct impact point is used. This is also not surprising
given that approximately $z \propto r^2$. Differences in the $z$
reconstruction arise through the modification of the relative position of the
tank with respect to shower core. It can be easily seen that when a tank gets
a closer position ($r$ decreases) as a result of the shift, the tanks placed
in the opposite $\zeta$ will get to a further one ($r$ increases).  This
suggests that the width of the reconstructed distribution should have a local
minimum when the correct impact point is considered.  An example is given
relaxing the $r$-cut to 500 m in Figs.~\ref{RMS60_19core_500}.

In a typical air shower array the core position is obtained by the relative
amplitude of the signals at different locations either through a fit or by
some other means. As for shifts in angular direction the minimum of the RMS
width of the $z$-distribution suggests that this method could be used either 
to reconstruct shower impact points independently or to check that the
core position reconstruction obtained through conventional methods is
consistent with the arrival time of the muons at large distances from shower
core, on an individual shower basis.

\section{Summary}
We have developed a method that has the potential of reconstructing the 
production altitude
for the muons in inclined cosmic ray showers based on the time distribution of
the muon signals in the detectors of an extensive air shower array. This
method requires knowledge of the arrival times of muons in the detectors of
the air shower array and it can be applied provided that a cut is made in
distance to shower axis, $r>r_c$.  Since the muon signal dominates at high $r$ 
it can be also used when the detectors cannot separate the muon signal provided
that the $r$ cut is chosen so that the muon signal dominates.

The method relies mainly on geometrical arguments and there are minor effects
introduced through the kinematical delay of the muons which have little effect
at large distances from shower axis.  The model does not rely on any
assumption about the interaction model for hadrons. Different models would give
rise to different kinematic corrections, but the effect is small. Although we
have assumed proton showers to explore the viability of this method the method
can be also used for heavier nuclei, and similar results would be obtained in
that case.  The necessary cut introduces limitations because of statistics.
We have checked that our method correctly reproduces the depth
distribution of muon
production using sets of simulated showers with AIRES to a degree of accuracy
that is zenith angle dependent.  The method works best in the
$60^\circ-80^\circ$ region and it is fairly stable with respect to
misreconstruction of the shower core and the incoming direction, in what
concerns the mean of the distribution. The RMS width of the production
distance distribution however is sensitive to both the reconstructed impact
point and arrival direction. The width displays a minimum when the correct
impact point and arrival direction are used in the reconstruction procedure.
 
This work represents a new approach to studying extensive air showers. It will
add information concerning the individual development of air showers and can
be used to check the reconstructed impact point and arrival directions. The
reconstruction of depth development in inclined showers can also have
important implications in improving the potential of air shower arrays to
detect neutrinos.

\section{Acknowledgments}
We thank J Alvarez--Mu\~niz and A.A.~Watson for many discussions on the time
structure of shower, and many helpful comments after reading the manuscript.
This work was partly supported by the Xunta de Galicia 
(PGIDIT02 PXIC 20611PN), by MCYT (FPA 2001-3837, FPA 2002-01161 and 
FPA 2004-01198). 
We thank CESGA, ``Centro de Supercomputaci\'on de Galicia'' for computer 
resources.

\appendix
\section{Modelling the distribution of surviving muons}

At ground level both muon energies and muon number are reduced because of
energy loss and decay. As a first approximation, assuming a constant energy
loss per unit depth $d E/dx=-\A$ along an uniform atmospheric density
$\rho$, both these effects can be easily accounted for.  After traveling a
distance $l$ a flux of muons $\phi_0$ of energy $E_i$ reduces by an energy
dependent factor to:
\begin{equation}
\phi(l) = \phi_0 \left[\frac{E_i-\rho \A l} {E_i} \right]^{\kappa}.
\label{N(l)}
\end{equation}
The last factor takes into account both energy loss and decay in flight.  The
muon mean lifetime, $\tau$, enters through the exponent $\kappa={mc^2/(\rho
\A c \tau)} \sim 0.8 $.  We can correct the energy of the produced muons 
given by Eq.~\ref{factorization} with such factor to obtain the distribution
in $E_i$, $p_t$, $z$ and $\zeta$ after traveling a distance $l$: 
\begin{equation}
\frac{d^4 N(l)}
{d z \;\! d\zeta \;\! d E_i \;\! d p_t} 
= {\cal N}_0 \;\! \frac{1}{2\pi}\;\! h(z) \;\! f_1(E_i) \;\! f_2(p_t) 
\left[\frac{E_i-\rho \A l}{E_i} \right]^{\kappa},
\label{factorization_N}
\end{equation}
Here $l$ is the distance traveled by the muon, which enters in Eq.~\ref{N(l)},
given by $l^2=(z-\Delta)^2 + r^2$, where $r$ is the distance to shower axis at
the end of the muon trajectory, and $z-\Delta$ is the distance travelled by
the muon measured along the shower axis. The correction $\Delta=r
\tan\theta\cos \zeta$ relates the distances measured along the axis between
the intercepts of shower axis and the muon trajectory with ground level and
depends on the zenith angle, $\theta$, as well as on the muon impact point
coordinates in the transverse plane, $r$ and $\zeta$.

For a muon to reach ground level there is a minimal production energy given 
by $E_i > m c^2+\rho \A l$. Typical values of $\rho \A l$ greatly exceed 
$m c^2$, particularly for inclined showers. 
After 
traveling a distance $l$ the transverse distance is simply $r=l \sin \alpha$.
Performing the change the coordinates from $p_t$ to $\sin \alpha$ in 
Eq.~\ref{factorization_N} we obtain: 
\begin{equation}
\frac{d^4 N(l)}{d z \;\! d\zeta \;\! d E_i
  \;\! d{\sin{\alpha}}} = {\cal N}_0 \;\! \frac{1}{2\pi}
\;\! h(z) \;\!  f_1(E_i)\;\! f_2\left(\frac{E_i}{c} \sin{\alpha} \right) \;\!
\left[1-\frac{\rho \A l}{E_i} \right]^{\kappa}\;\! \frac{E_i}{c}.
\label{mu-dist}
\end{equation}
Correlations 
between the ground variables appear naturally because of energy loss and
decay. 

We can now introduce the following parameterizations for $f_1(E_i)$
and $f_2(p_t)$ which were shown in \cite{cazon3} to give good approximations
to the muon time distributions:
\begin{eqnarray}
f_1(E_i)= \frac{\gamma -1}{ mc^2} 
\left(\frac{E_i}{mc^2}\right)^{-\gamma}\Theta(E_i-mc^2), \\
f_2(p_t)= \frac{p_t}{Q^2} \exp\left(-\frac{p_t}{Q}\right),
\end{eqnarray}
where $\gamma\simeq 2.6$ and $Q \simeq 0.17$ GeV/c.  Eq.~\ref{mu-dist} now
becomes
\begin{equation}
\frac{d^4 N}{d z \;\! d\zeta \;\! d E_i \;\! d{\sin{\alpha}}}
= \frac{{\cal N}_0 (\gamma -1) }{2\pi} h(z) \;\! 
\left(\frac{E_i}{mc^2}\right)^{-\gamma+1} 
\;\! \frac{E_i \sin\alpha}{c^2 Q^2} \exp\left(-\frac{E_i \sin\alpha}{c  Q}\right) 
\left[1-\frac{\rho \A l}{E_i} \right]^{\kappa}.
\end{equation}
We now integrate the distribution in $E_i$ for fixed $z$ and $\alpha$ to 
obtain: 
\begin{equation}
\frac{d^3 N}{d z \;\! d\zeta \;\! d{\sin{\alpha}}} = 
\int_{mc^2 + \rho\A l} ^{\infty} \frac{d^4 N}{d z \;\! d\zeta 
d E_i \;\! d{\sin{\alpha}}} \;\! d E_i =
{ \frac{{\cal N}_0 (\gamma -1)}{ 2\pi}} \left( 
\frac{mc^2}{Q c}\right)^{\gamma -1} 
h(z) \;\! I(l,\sin \alpha) \sin^{\gamma -2} \alpha,
\label{SinDistribution}
\end{equation}
in which $I(l,\sin \alpha)$ is a dimensionless integral in the variable $x=
\frac{E_i \sin \alpha}{c Q}$:
\begin{equation}
I(l,\sin \alpha)= \int_{y + x_0} ^{\infty}  x^{-\gamma+2} 
\left[1-\frac{x_0}{x}\right]^{\kappa} \exp \left(-x\right) dx,
\end{equation}
with $x_0(l,\alpha)=\frac{\rho \A l}{cQ} \sin \alpha$ and
$y(\alpha)=\frac{mc^2}{cQ}\sin\alpha$.  We note that $y\ll x_0$ when $l \gg
mc^2/\rho \A \simeq 500$~m and in that case the integral can be approximated
replacing its lower limit by $x_0$.  Since $z$, the distance of muon
production, is relatively large, particularly for inclined showers, this
approximation is adequate for most circumstances. Then it is easily seen that
the integral depends only on the combination $l \sin \alpha=r$, i.e. on the
transverse distance.  We can then replace $I(l,\sin \alpha)$ by $I(r)$.

In terms of the differential solid angle $d^2 \Omega=-d\zeta d \cos \alpha$ 
the number of muons arriving at ground level coming from production distance 
$z$ becomes:
\begin{equation}
\frac{d^3 N(r)}{ d z \;\! d^2\Omega} = 
-\frac{d^3 N(r)}{ d z \;\! d \zeta \;\! d\cos \alpha} = 
\frac{{\cal N}_0 (\gamma -1)}{2\pi} \left(\frac{mc^2}{ Q c}\right)^{\gamma -1} 
h(z) ~ I(r) 
\left( \frac{ \cos \alpha }{\sin \alpha} \right) \sin^{\gamma-2} \alpha.
\label{CosDistribution}
\end{equation}

%
\begin{figure}[htb]
\begin{center}
\includegraphics[width=16 cm,clip]{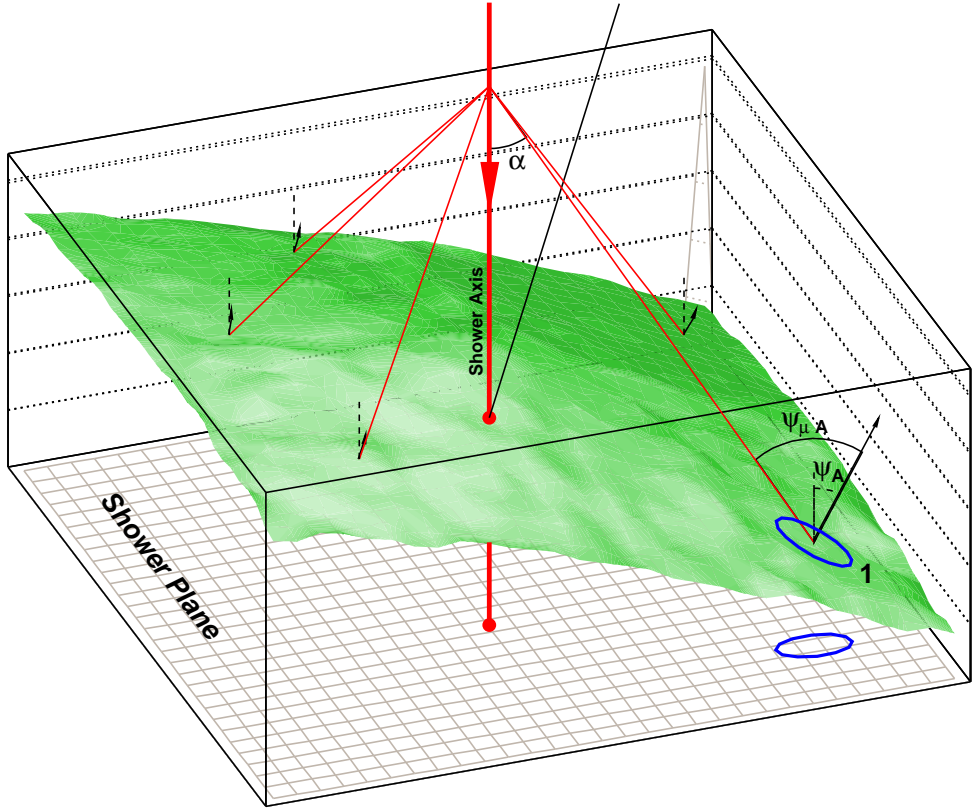}
\caption{
Illustration of the relation between the ground surface, the transverse plane
(shower plane) and the shower axis (big arrow). The small arrows represent a
detector surface direction normal to the detector surface, the
dashed lines are parallel to shower axis and the continuous lines
are the muon trajectories.  The expressions for number of
muons $dN_A$ which go through the surface element $d^2A$ (labeled 1) are
described in the text.} 
\label{f:terra2}
\end{center} 
\end{figure}
Using the relations between $z$, $l$ and $\alpha$, we can substitute
Eq.~\ref{CosDistribution} into Eq.~\ref{rho_mu} and integrate in $z$ to
obtain:
\begin{equation}
\hat{N}_{Ar \zeta} \equiv \frac{1}{r} \;\! \frac{d^2N_A}{dr \;\! d \zeta}= 
\frac{ {\cal N}_0 (\gamma -1) }{2\pi} 
\left(\frac{mc^2}{Q c}\right)^{\gamma -1} 
 ~ \frac{I(r)}{r}~ \int h(z)
\;\! \frac{ \cos \alpha ~ D_A(\Omega)  \sin^{\gamma-2}\alpha}{l}  d z.
\label{N_r}  
\end{equation}
$\hat{N}_{Ar \zeta}$ is the number of muons going through $d^2 A$ (whose
projection onto the transverse plane is $r dr d\zeta$) and has a non-trivial
geometrical dependence through the $D_A(\Omega)$ factor.  This factor can be
both greater or less than 1 and becomes 1 as the angle $\alpha$ tends to
zero, that is in the limit of small $r$.  In that case $\hat{N}_{Ar \zeta}$
simply becomes the muon surface density in the transverse plane.
$D_A(\Omega)$ can be regarded as a geometrical correction which accounts for
the fact that the muons do not travel parallel to the shower axis and the fact
that the detector planes are not perpendicular to the shower axis not to the
muon fluxes (See Fig.~\ref{f:terra2}). 

It can be shown that when $z$ and $r$ are fixed $D_A$ depends on the 
angle $\zeta$ because the angles $\alpha$ and $\psi_{\mu A}$ differ. 
As a result the factor $D_A$ is responsible for part of the asymmetry in the 
muon signal. For instance a horizontal surface will have larger efficiency 
for collecting early arriving muons because $\psi_{\mu A}$ is smaller than 
for late arriving muons. Note that $D_A(\Omega)$ would be exactly 1 for 
an ``ideal'' detector of spherical shape. In that case these differences  
disappear. 

The expression obtained in Eq.~\ref{g(t)2} relates the time distribution to a
transform of the depth distribution $h(z)$ which effectively accounts for 
muon decay in flight through $l$. If we are interested in
obtaining the production distribution $h(z)$ formally we can rewrite 
Eq.~\ref{g(t)2} as: 
\begin{equation}
h(z) =  g(t) \frac{dt}{dz} l^{-1+\gamma} \cos^{-1} \alpha ~ 
D_A^{-1}(\Omega) \int_0^{\infty} h(z) \;\! 
{l}^{1-\gamma} \cos \alpha ~D_A(\Omega) dz.
\label{h_z}
\end{equation} 
This expression in principle allows us to obtain $h(z)$ from the time
distribution of the arriving muons at a given point on the ground, $g(t)$. On
its own it is not very useful because typically a single detector in an air 
shower array does not collect sufficient statistics to sample $h(z)$
reliably. in a practical situation one must combine the results of several 
detectors. Since $h(z)$ is normalized to 1, the unknown factor which is given 
by the integral on the RHS of the equation acts as an effective weight to be 
given to each detector. In a first attempt it is possible to ignore it. 
For inclined showers the weights to be applied for the relevant detectors 
are expected to be quite similar and the approximation works well. 
More sophisticated approaches could be deviced for instance using 
Eq.~\ref{h_z} with a trial $h(z)$ function in an iterative
process to sample $h(z)$. However since $h(z)$ is not directly available from
the simulation program used we will not need to calculate $h(z)$ and we have  
instead compared our results to the $z$-distribution of the surviving muons.

\section{Parameterization of kinematical delays}

The mean kinematical time delay can be obtained by applying the method
developed in Ref.~\cite{cazon3}, summarized in Eq~\ref{t:kinetic} to the 
distributions discussed in this article:
\begin{equation}
< t_{\varepsilon}  > \  = 
\frac{\int  t_{\varepsilon} \frac{d^4 N}{dz \;\! d\zeta \;\! dE_i \;\! dr  } d E_i
    }{\int   \frac{d^4 N}{dz \;\! d\zeta \;\! dE_i \;\! dr} dE_i }.
\end{equation}
After some manipulation, using the results of the models in Appendix A  
a simple expression can be obtained for it: 
\begin{equation}
< t_{\varepsilon}  > \ = 
\ \frac{1}{2c}\frac{r^2}{l}  {\left(\frac{mc^2}{cQ}\right)^2} \frac{
\int_{y + x_0} ^{\infty}  x^{-\gamma} 
\left[1-\frac{x_0}{x}\right]^{\kappa-1} \exp \left(-x\right) dx}{
\int_{y + x_0} ^{\infty}  x^{-\gamma+2} 
\left[1-\frac{x_0}{x}\right]^{\kappa} \exp \left(-x\right) dx } = 
\frac{1}{2c}\frac{r^2}{l} \epsilon(r,z-\Delta),
\label{eq:tepsilon_r}
\end{equation}
with $x_0(l,\alpha)=\frac{\rho \A l}{cQ} \sin \alpha$ and
$y(\alpha)=\frac{mc^2}{cQ}\sin\alpha$.

In the last equality of the above expressions we have introduced the
dimensionless factor $\epsilon(r,z-\Delta)$. We note that for $z-\Delta \gg
r$, for instance in inclined showers, it gives the ratio of the average
kinematical delay to the geometrical delay at a given position
$\frac{<t_\varepsilon>}{t_g}\simeq \epsilon(r,z-\Delta)$. This indicates the
regions where the geometric delay dominates, which depend on production
distance.
For practical purposes we have parameterized $\epsilon(r,z)$ as follows:
\begin{equation}
\epsilon (r,z) = p_0(z)  \left(\frac{r}{\rm m}\right)^{p_1},
\end{equation}
with 
\begin{eqnarray}
\log_{10}p_0(z) &=& -0.6085   +1.955 \ \log_{10} (z/m)
                    -0.3299 \ \log^2_{10} (z/m) +0.0186 \ 
		    \log^3_{10} (z/m), \\
\log_{10}p_1    &=& -1.176.
\end{eqnarray}
%


\end{document}